\documentclass[final,5p,times,twocolumn]{elsarticle}

\usepackage{graphicx, epsfig}
\usepackage{multirow}
\usepackage{color}
  
\usepackage{amssymb}

\journal{Physics Letters B}

\begin{document}

\begin{frontmatter}



\title{Kaon and Pion Production in Central Au+Au Collisions at $\sqrt{s_{NN}}=62.4$~GeV\tnoteref{t1}}
\tnotetext[t1]{The BRAHMS Collaboration}

\author[oslo]{I.C.Arsene\corref{cor1}}
\author[nbi]{I.G.Bearden}
\author[bnl]{D.Beavis}
\author[uk]{S.Bekele\fnref{fn1}}
\author[buca]{ C.Besliu}
\author[nyu]{B.Budick}
\author[nbi]{H.B{\o}ggild}
\author[bnl]{C.Chasman}
\author[nbi]{C.H.Christensen}
\author[nbi]{P.Christiansen\fnref{dddag}}
\author[nbi]{H.H.Dalsgaard}
\author[bnl]{R.Debbe}
\author[nbi]{{J.J.Gaardh{\o}je}}
\author[tamu]{ K.Hagel}
\author[bnl]{ H.Ito}
\author[buca]{ A.Jipa}
\author[uk]{E.B.Johnson\fnref{dag}}
\author[nbi]{ C.E.J{\o}rgensen\fnref{dddddag}}
\author[krakow]{ R.Karabowicz}
\author[krakow]{ N.Katrynska}
\author[uk]{E.J.Kim\fnref{ddag}}
\author[nbi]{ T.M.Larsen}
\author[bnl]{ J.H.Lee}
\author[oslo]{ G.L{\o}vh{\o}iden}
\author[krakow]{ Z.Majka}
\author[uk]{ M.J.Murray}
\author[tamu]{ J.Natowitz}
\author[nbi]{ B.S.Nielsen}
\author[nbi]{ C.Nygaard}
\author[uk]{ D.Pal}
\author[oslo]{A.Qviller}
\author[iphc]{F.Rami}
\author[nbi]{ C.Ristea}
\author[buca]{ O.Ristea}
\author[uib]{ D.R{\"o}hrich}
\author[uk]{ S.J.Sanders}
\author[krakow]{ P.Staszel}
\author[oslo]{ T.S.Tveter}
\author[bnl]{F.Videb{\ae}k\fnref{sp}}
\author[tamu]{ R.Wada}
\author[uib]{ H.Yang\fnref{ddddddag}}
\author[uib]{Z.Yin\fnref{ddddag}}
\author[ss]{and I.S.Zgura}

\address[bnl]{Brookhaven National Laboratory, Upton, New York, USA}
\address[iphc]{Institut Pluridisciplinaire Hubert Curien et Universit{\'e} de Strasbourg, 
Strasbourg, France}
\address[ss]{Institute of Space Science,Bucharest-Magurele, Romania}
\address[krakow]{M. Smoluchowski Inst. of Physics,  Jagiellonian University, Krakow, Poland}
\address[nyu]{New York University, New  York, USA,}
\address[nbi]{Niels Bohr Institute,  University of Copenhagen, Copenhagen, Denmark}
\address[tamu]{Texas A$\&$M University, College Station, Texas, USA}
\address[uib]{University of Bergen, Department of Physics and Technology, Bergen,  Norway}
\address[buca]{University of Bucharest, Romania}
\address[uk]{University of Kansas, Lawrence, Kansas, USA }
\address[oslo]{University of Oslo, Department of Physics, Oslo, Norway}

\fntext[fn1]{Present Address: Dept of Physics, Tennessee Tech University,Cookeville, Tennessee}
\fntext[dag]{Present address: Radiation Monitoring Devices, Cambridge, MA, USA}
\fntext[ddag]{ Present address: Division of Science Education, Chonbuk National University, Jeonju, Korea }
\fntext[dddag]{Present Address: Div. of Experimental
High-Energy Physics, Lund University, Lund, Sweden}
\fntext[ddddag]{Present Address: Institute of Particle Physics, Huazhong Normal University,Wuhan,China}
\fntext[dddddag]{Present Address: Ris{\o} National Laboratory, Denmark}
\fntext[ddddddag]{Present Address: Physics Institute, University of Heidelberg, Heidelberg, Germany}
\fntext[sp]{Spokesperson \it{e-mail: videbaek@bnl.gov}}
\cortext[cor1]{Corresponding Author. {\it e-mail: i.c.arsene@fys.uio.no}(I.C.Arsene) \\
Present Address: ExtreMe Matter Institute EMMI, GSI Helmholtzzentrum f{\"u}r Schwerionenforschung GmBH,
Darmstadt, Germany}

\begin{abstract}
Invariant $p_T$ spectra and rapidity densities covering a large rapidity range ($-0.1<y<3.5$) 
are presented for $\pi^{\pm}$ and $K^{\pm}$ mesons from central Au+Au collisions at $\sqrt{s_{NN}}=62.4$~GeV.
The mid-rapidity yields of meson particles relative to their anti-particles are found to be close to unity
($\pi^{-}/\pi^{+} \sim 1$, $K^{-}/K^{+} \sim 0.85$) while the anti-proton to proton ratio is
$\bar{p}/p \sim 0.49$. The rapidity dependence of the $\pi^{-}/\pi^{+}$ ratio is consistent with
a small increase towards forward rapidities while the $K^{-}/K^{+}$ and $\bar{p}/p$ ratios show a steep decrease
to $\sim 0.3$ for kaons and $0.022$ for protons at $y \sim 3$.
It is observed that the kaon production relative to its own anti-particle as well as to pion  
production in wide 
rapidity and energy ranges shows an apparent universal behavior  
consistent with the baryo-chemical potential, as deduced from the $\bar{p}/p$ ratio, being the driving parameter.
\end{abstract}


\end{frontmatter}


\section*{Introduction}

As the collision energy has increased with the advent of new relativistic heavy-ion
accelerators, from AGS energies ($\sqrt{s_{NN}} \leq 4.9$~GeV) to those achieved with the 
SPS ($\sqrt{s_{NN}} \leq 17.3$~GeV) and recently with RHIC ($\sqrt{s_{NN}} \leq 200$~GeV),
the system created in heavy-ion collisions has been found to
evolve from one that is baryon rich to one dominated by  mesons 
\cite{AGSpion, AGSkaon, SPSmeson, BRAHMSmeson}.  
This change is evident in the increase of
central rapidity densities of emitted pions and kaons accompanied by the simultaneous shift of 
the net-baryon peak one or two units of rapidity down from beam rapidity.
Due to this shift, the net-baryon peak gradually moves from mid-rapidity (AGS and SPS)
\cite{AGSproton, SPSproton} towards high rapidity (RHIC) \cite{BRAHMSproton},
leaving a relatively net-baryon poor region at mid-rapidity at the highest RHIC energy.

At AGS energies, the observed ratio of strange particles to
pions in A+A collisions is larger than 
that measured in either p+p or p+A 
collisions and this ratio increases significantly 
with beam energy \cite{AGSkaon, AGSkpiEnergy, AGSlambda}. 
This behavior is understood within cascade models as 
arising from hadronic rescatterings involving heavy baryon resonances \cite{RQMD1, RQMD2}.
In the SPS energy range it is conjectured \cite{SPSprediction} 
that the nuclear fireball undergoes a phase transition
from bound hadronic matter to a deconfined quark-gluon plasma state (QGP). 
This conjecture is supported by the observation that excitation functions of the $K^{+}/\pi^{+}$, 
$\Lambda/\pi$, and $\Xi^{-}/\pi$ ratios at mid-rapidity
are found to peak around $\sqrt{s_{NN}} = 7.6$~GeV \cite{SPSkpiEnergy} and then decrease slightly with 
increasing energy. 
The energy dependence of other observables, like the onset of a plateau for the inverse
slope parameter of kaon spectra
and a kink in the $4\pi$ pion yields normalized to the number of participant nucleons $N_{\pi} / N_{part}$,
is also used to support the transition to a deconfined phase at $\sqrt{s_{NN}} = 7.6$~GeV \cite{SPSkpiEnergy}.
The gross features of the $K/\pi$  excitation function can be described 
in theoretical models, but additional features are needed to describe the data quantitatively.
Phenomenologically motivated thermal models 
can explain the observed behavior with the assumption of a QGP formed in the early stages 
of the collision \cite{SPSprediction}.
However, more recent thermal models are able to reproduce the 'horn' structure of the
excitation function by including higher hadronic resonant states without the assumption
of a phase transition \cite{andronic1}.
The $K^{-}/\pi^{-}$ ratio has a monotonically increasing energy dependence and is always
lower than the $K^{+}/\pi^{+}$ ratio because of the larger share of strange quarks,
which form baryons (mainly $\Lambda$'s) \cite{SPSkpiEnergy}.
From the top SPS energy to the top RHIC energy the 
$K^{+}/\pi^{+}$ ratio
at mid-rapidity is fairly constant,
suggesting that the nuclear medium reaches chemical equilibrium with respect to the production
of strange quarks \cite{BRAHMSmeson}. 

\begin{table*}[thp]
\caption{$dN/dy$ values for charged pions as a function of rapidity. The given errors are statistical only.
The $p_T$ range specified represents the fit range.
Fiducial yield is the integrated yield covered by the data.}
\centering
\begin{tabular}{c|ccc|ccc|cc}
\hline
\multirow{2}{*}{y} & \multicolumn{3}{|c|}{$dN/dy$ power law} & \multicolumn{3}{|c|}{$dN/dy$ $m_T$ expo} &
\multicolumn{2}{|c}{fiducial yields} \\
& $\pi^{+}$ & $\pi^{-}$ & $p_T$ [GeV/$c$] & $\pi^{+}$ & $\pi^{-}$ & $p_T$ [GeV/$c$] & $\pi^{+}$ & $\pi^{-}$ \\
\hline
-0.20$\leq$y$\leq$0.00 & 224.0$\pm$2.7 & 232.4$\pm$2.9 & 0.20 - 2.00 & 195.8$\pm$2.2 & 202.1$\pm$2.4 & 0.20 - 1.00 & 
    167.9$\pm$2.1 & 170.4$\pm$2.3 \\
 0.00$\leq$y$\leq$0.20 & 231.9$\pm$3.2 & 233.7$\pm$2.9 & 0.20 - 2.00 & 202.7$\pm$2.7 & 203.6$\pm$2.3 & 0.20 - 1.00 &
    171.7$\pm$2.5 & 171.7$\pm$2.1 \\
 0.70$\leq$y$\leq$0.90 & 209.7$\pm$2.1 & 213.1$\pm$2.2 & 0.35 - 1.90 & 180.7$\pm$1.3 & 182.4$\pm$1.4 & 0.35 - 1.00 & 
     99.7$\pm$0.6 & 100.3$\pm$0.6 \\
 0.90$\leq$y$\leq$1.10 & 205.5$\pm$1.9 & 214.3$\pm$1.9 & 0.20 - 1.80 & 178.6$\pm$1.2 & 185.4$\pm$1.1 & 0.20 - 1.00 &
    151.1$\pm$1.7 & 135.9$\pm$0.9 \\
 3.05$\leq$y$\leq$3.15 &  57.1$\pm$1.9 &  68.6$\pm$3.2 & 0.25 - 1.60 &  49.9$\pm$0.9 &  52.2$\pm$0.8 & 0.25 - 1.60 &
     22.4$\pm$0.5 &  24.8$\pm$0.5 \\
 3.15$\leq$y$\leq$3.25 &  49.2$\pm$1.4 &  60.5$\pm$2.0 & 0.25 - 1.50 &  45.8$\pm$1.1 &  53.1$\pm$1.2 & 0.25 - 1.50 &
     11.0$\pm$0.4 &   8.3$\pm$0.2 \\
 3.40$\leq$y$\leq$3.60 &  24.8$\pm$2.0 &  24.3$\pm$3.7 & 0.50 - 1.00 &  20.3$\pm$1.2 &  19.9$\pm$1.0 & 0.50 - 1.00 &
      4.8$\pm$0.1 &   4.9$\pm$0.1 \\
\hline
\end{tabular}
\label{table1}
\end{table*}

We present invariant $p_T$ spectra of  pions and kaons in central ($0-10\%$)  Au+Au collisions at
$\sqrt{s_{NN}}=62.4$~GeV measured by the BRAHMS spectrometers\cite{BRAHMSwhite}.  
At this energy it is possible by varying the spectrometer angles to 
achieve an overlap in terms of the baryo-chemical potential (as reflected in 
the $\bar{p}/p$ ratio) of the RHIC and lower energy SPS results.
The data cover the rapidity interval $-0.1 < y < 3.5$, 
which comes close to the beam rapidity ($y_{beam} = 4.2$).   
Total yields deduced from the spectra are used to  
calculate particle ratios that can be compared to 
the lower energy SPS results and to theoretical models. 
The large rapidity coverage allows very different nuclear media to be probed: 
at mid-rapidity the fireball is dense 
and the yields of produced anti-hadrons and hadrons are similar
($\pi^{-}/\pi^{+} \sim 1$, $K^{-}/K^{+} \sim 0.85$, $\bar{p}/p \sim 0.45$), while at 
forward rapidity, in the fragmentation region, we observe a net-baryon rich medium 
with 4 times smaller 
meson densities and a 20 times smaller $\bar{p}/p$ ratio.
By covering a large interval for important physical parameters ({\it e.g.} rapidity, particle density, net-baryon density,
baryo-chemical potential), these measurements will provide additional constraints to
those models that support the conjecture of a phase transition to quark-gluon plasma.

\begin{table*}[thp]
\caption{$dN/dy$ values for charged kaons as a function of rapidity. The errors are statistical only. 
The $p_T$ range specified represents the fit range.
Fiducial yield is the integrated yield covered by the data.}
\centering
\begin{tabular}{c|ccc|ccc|cc}
\hline
\multirow{2}{*}{y} & \multicolumn{3}{|c|}{$dN/dy$ $m_T$ expo} & \multicolumn{3}{|c|}{$dN/dy$ Boltzmann} &
\multicolumn{2}{|c}{fiducial yields} \\
& $K^{+}$ & $K^{-}$ & $p_T$ [GeV/$c$] & $K^{+}$ & $K^{-}$ & $p_T$ [GeV/$c$] & $K^{+}$ & $K^{-}$ \\
\hline
-0.15$\leq$y$\leq$0.15 & 35.6$\pm$0.9 & 30.4$\pm$0.8 & 0.45 - 1.90 & 33.6$\pm$0.8 & 28.6$\pm$0.8 & 0.45 - 1.90 &
                 19.5$\pm$0.5 & 16.6$\pm$0.5 \\
 0.60$\leq$y$\leq$0.80 & 33.4$\pm$0.4 & 27.5$\pm$0.4 & 0.35 - 1.80 & 31.6$\pm$0.4 & 26.0$\pm$0.4 & 0.35 - 1.90 &
                 23.9$\pm$0.3 & 15.0$\pm$0.2 \\
 0.80$\leq$y$\leq$1.00 & 32.2$\pm$0.6 & 26.7$\pm$0.5 & 0.40 - 1.80 & 30.1$\pm$0.5 & 26.7$\pm$0.5 & 0.40 - 1.80 &
                 19.1$\pm$0.4 & 11.7$\pm$0.2 \\
 2.55$\leq$y$\leq$2.65 & 15.2$\pm$0.4 &  8.8$\pm$0.3 & 0.25 - 1.30 & 14.8$\pm$0.4 &  8.6$\pm$0.3 & 0.25 - 1.30 &
                 11.5$\pm$0.4 &  6.6$\pm$0.2 \\
 2.65$\leq$y$\leq$2.75 & 14.0$\pm$0.4 &  8.2$\pm$0.3 & 0.35 - 1.20 & 13.5$\pm$0.4 &  7.9$\pm$0.2 & 0.35 - 1.20 &
                  6.6$\pm$0.2 &  4.7$\pm$0.14 \\
 3.15$\leq$y$\leq$3.25 &  8.1$\pm$0.6 &  2.9$\pm$0.3 & 0.60 - 1.50 &  7.5$\pm$0.5 &  2.7$\pm$0.3 & 0.60 - 1.50 & 
                  2.44$\pm$0.12 &  0.84$\pm$0.06 \\
 3.25$\leq$y$\leq$3.35 &  6.8$\pm$0.3 &  2.34$\pm$0.14 & 0.50 - 1.20 &  6.3$\pm$0.3 &  2.20$\pm$0.14 & 0.50 - 1.20 &
                  1.83$\pm$0.07 &  0.75$\pm$0.04 \\
\hline
\end{tabular}
\label{table2}
\end{table*}

\section*{Experimental Setup and Data Analysis}
The BRAHMS experiment consists of global event characterization 
detectors and two spectrometer arms. Collision centrality is determined using a hybrid
array of Si strip detectors and plastic scintillator tiles located at the nominal 
interaction point.   
The Mid-Rapidity Spectrometer (MRS) and the Forward Spectrometer (FS) 
are small solid angle spectrometers that can be rotated in the horizontal plane so as together be able to achieve
polar angular coverage from 90$^{\circ}$ to 2.3$^{\circ}$. 
\begin{figure}[htp]
\begin{center}
\includegraphics[width=1.0\columnwidth]{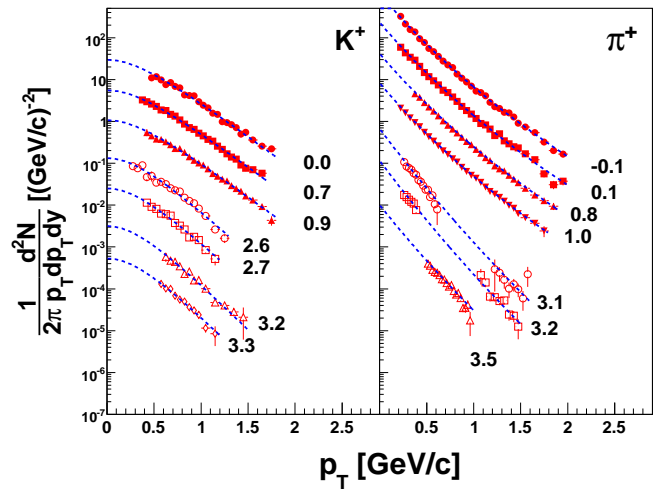}
\caption{(Color online) Spectra for $K^{+}$ (left) and $\pi^{+}$ (right) from 0-10\% central Au+Au collisions at $\sqrt{s_{NN}}=62.4$~GeV 
in selected rapidity slices. 
Each spectrum is multiplied with a factor of $0.2^n$ for better visibility.
For the kaon spectra, the values of $n$ are 0,1,2,3,4,5,6 for the rapidities $y$ = 0.0,0.7,0.9,2.6,2.7,3.2 and 3.3, 
respectively. For the pion spectra, the values of $n$ are 0,1,2,3,4,5,6 for the rapidities 
$y$ = -0.1,0.1,0.8,1.0,3.1,3.2 and 3.5, respectively.
The fits are $m_T$ exponentials for kaons and $p_T$ power 
laws for pions. The numbers on the plot indicate the rapidity corresponding to each spectrum. 
The error bars represent statistical errors.}
\label{spectra}
\end{center}
\end{figure}
For the present studies,
the MRS spectrometer was positioned at 90$^{\circ}$, 45$^{\circ}$ and 40$^{\circ}$ and
the FS at 6$^{\circ}$, 4$^{\circ}$ and 3$^{\circ}$ at a few magnetic field settings,
due to the limited running time.

Particle identification (PID) is achieved in both spectrometers using time-of-flight walls
(TOFW in MRS and H2 in FS). In the FS a ring imaging Cherenkov detector (RICH) is used in addition. 
A detailed description of the technical capabilities
of the experimental setup can be found in references \cite{BRAHMSnim} and
\cite{BRAHMSrich}.
The identification of charged pions and kaons with the time-of-flight detectors was done by using cuts in ($m^2, p$) space, 
where $m^2 = p^2(t^2_{TOF}/L^2 - 1/c^2)$ is determined for a given momentum $p$ by using
the time of flight ($t_{TOF}$) and the track path length($L$). 
Momentum dependent three-$\sigma$ cuts about the mean $m^2$ of a given species were applied 
with the additional condition that the level of
contamination from other species at a given $m^2$ and $p$ is limited to 5\% inside the cut. 
Clean $\pi-K$ separation is achieved up to momenta of 1.5 GeV/$c$
in the MRS and 3 GeV/$c$ in the FS.  For higher momenta the 3$\sigma$ 
curves overlap and the contamination condition gives asymmetrical cuts in the $(m^2, p)$ 
space, necessitating the use of momentum dependent cuts and corrections. For the TOFW, the
momentum dependent PID corrections ranges from 0 at 1.5 GeV/$c$ up to 20\% for pions and 50\% 
for kaons at 2.2 GeV/$c$. In H2, the same correction ranges from 0 at 3 GeV/$c$ 
to 10\% and 80\% at 4.5 GeV/$c$ for pions and kaons, respectively.
In the FS, the RICH was used to identify pions and kaons at higher momenta. 
The pions are separated from kaons by comparing the  reconstructed Cherenkov ring radii for momenta with the calculated value for pions in the momentum range of 2.5 to 20 GeV/$c$. 
Kaons are identified by the RICH above 10 GeV/$c$ where the RICH efficiency saturates at a value of $97\%$ \cite{BRAHMSrich}. 
Between 4.5 GeV/$c$ and 9 GeV/$c$ an 
indirect method labels as kaon particles those that give no signal in the RICH but are also
not identified as protons in H2. The contamination of kaons with pions unresolved in RICH 
and protons unresolved in H2 depends on momentum and on relative particle abundances. 
For $K^{-}$ the contaminant contribution is estimated to be almost constant at $20\%$ 
whereas for $K^{+}$ it varies from
$14\%$ to $40\%$. 

Invariant differential yields, $\frac{1}{2\pi}\frac{d^2N}{p_T dp_T dy}$, were constructed
for each spectrometer setting and were corrected for geometrical
acceptance, tracking and PID efficiency, contamination, 
in-flight weak decays and multiple scattering effects by using a Monte Carlo calculation
simulating the geometry and tracking of the BRAHMS detector system.
The feed-down correction for charged pions originating from the weak decays of
$K_{S}^{0}$ and $\Lambda$ was applied as described in reference \cite{BRAHMSmeson}
and amounts to $5\%$ of the measured yield in MRS settings and $7\%$ in FS settings.
The feed-down correction was applied directly to the $p_{T}$ integrated rapidity densities.
By merging all of the spectrometer magnet settings, 
invariant $p_T$ spectra were extracted in 
several rapidity intervals and fitted with different
functions in order to extract the integrated yields (see Fig.~\ref{spectra}). 
We estimate the point-to-point systematic errors to be $\leq$5\% and the
systematic errors from normalization, tracking efficiencies and other corrections to be $\sim$8\%.
The spectrometer acceptances allow measurements down to 
$p_T=0.2$~GeV/$c$
for pions and  $p_T=0.25$~GeV/$c$
for kaons, with the absence of lower momentum
data contributing to the systematic errors for the integral yield. 
This uncertainty is largest for pions because of their lowest average $p_T$.
In order to estimate the extrapolation uncertainty,  
we used both a power law distribution of the 
form $A(1+p_T/p_0)^{-B}$ and an $m_T$ exponential function to fit the pion spectra. 
Of these two, the power law distribution gives the best fit to the experimental
results over the observed momentum range, and for very low $p_T<0.1$~GeV/$c$ agrees 
with the measurements made by the PHOBOS collaboration  at $y=0.8$ \cite{PHOBOS62}.  
For the $m_T$ exponential function the fit range is
limited to $p_T < 1$~GeV/$c$.
The kaon spectra are equally well described 
by an $m_T$ exponential function, $Ae^{(-m_T/T)}$, and by a  Boltzmann distribution.
All the presented data are tabulated and available on the BRAHMS website \cite{brahmsWeb}.

\begin{figure}[htp]
\begin{center}
\includegraphics[width=1.0\columnwidth]{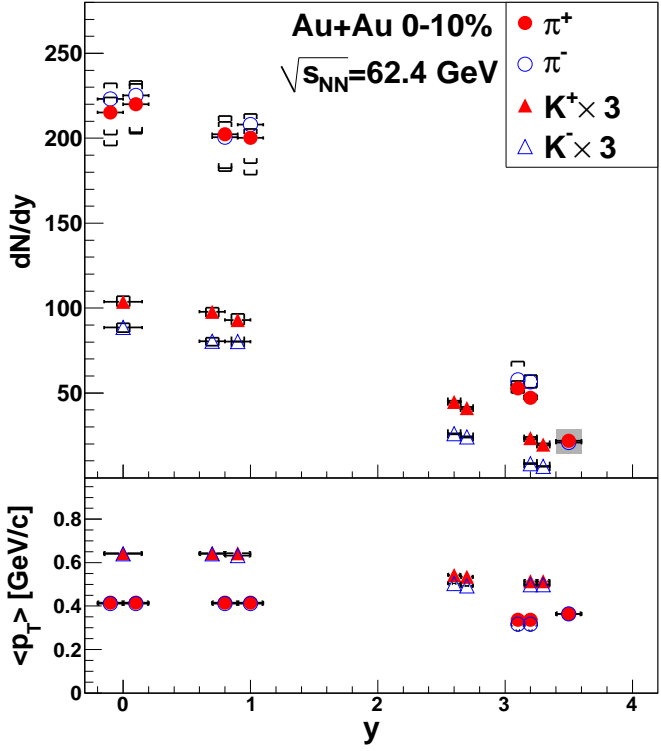}
\caption{(Color online) Upper panel: $dN/dy$ as a function of rapidity for $\pi^{\pm}$ and $K^{\pm}$ from 0-10\% central
Au+Au collisions at $\sqrt{s_{NN}}=62.4$~GeV. 
The error bars are statistical. The square brackets below and above each data point indicate the 
$dN/dy$ values extracted with the selected functionals (see Tables \ref{table1} and \ref{table2} and text).
The experimental uncertainties on the extrapolated yields are below 5\% and not shown in the figure, 
except for the pion yields at $y=3.5$ where these amount to $\sim30$\% and are indicated by the gray box.
Bottom panel: $\langle p_T\rangle$ dependence on $y$. The statistical error bars are covered by the symbols. 
}
\label{dndy}
\end{center}
\end{figure}

\section*{Results and Discussion}
The resulting $p_T$ integrated yields, $dN/dy$, for pion and kaons are shown in Tables \ref{table1} and \ref{table2}.  
The $dN/dy$ distributions, taking the statistically weighted 
average of the two functional forms used in fitting each species, 
are shown in the 
upper panel of Fig.~\ref{dndy}. 
The horizontal error bars indicate the width of the rapidity slices and the vertical error bars, 
smaller than the marker size, are statistical errors. The square brackets below and above each
data point indicate the $dN/dy$ values extracted with the employed functionals (see Tables \ref{table1} and \ref{table2})
and serve as an estimate for the systematic errors due to extrapolation to low transverse momentum.

The bottom panel of Fig.~\ref{dndy} shows the rapidity dependence of the average
transverse momentum $\langle p_T\rangle$ for pions
and kaons. For pions, $\left<p_T\right> \sim 0.41$~GeV/$c$ at $y=0$ and decreases to $\sim0.32$~GeV/$c$ at $y>3.0$ while 
for kaons, $\left<p_T\right>$ drops from $\sim0.65$~GeV/$c$ at $y=0$ to $\sim0.5$~GeV/$c$ at $y=3.2$. 
The drop of the averaged transverse momentum at forward rapidity is accompanied by a 
relatively large decrease in particle densities and in collective models 
can be explained by the decrease of the collective radial flow velocity.

\begin{figure}[htp]
\begin{center}
\includegraphics[width=1.0\columnwidth]{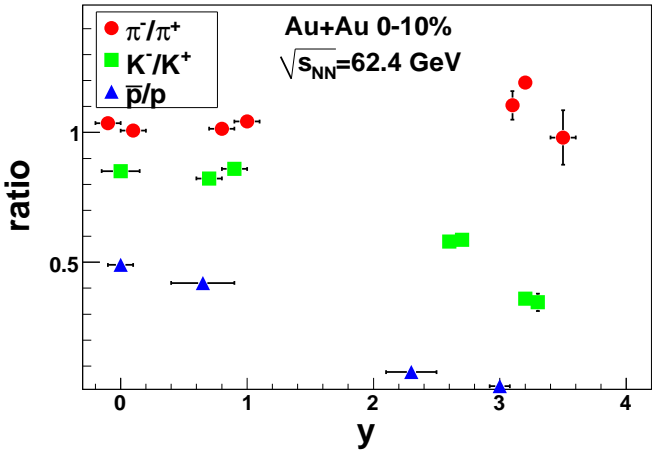}
\caption{(Color online) Anti-particle to particle ratios as a function of rapidity in 0-10\% central Au+Au collisions at $\sqrt{s_{NN}}=62.4$~GeV. 
The error bars are statistical only. 
}
\label{likeRatios}
\end{center}
\end{figure}

Figure ~\ref{likeRatios} shows the rapidity-dependent
anti-hadron to hadron integrated yield ratios for pions, kaons and protons.
The $\bar{p}/p$ ratios are obtained by using the yields presented in reference \cite{BRAHMS62proton}.
The baryon yields are not corrected for feed-down from hyperons (see \cite{BRAHMS62proton}).
The $\pi^{-}/\pi^{+}$ ratio is approximately equal to unity over the entire rapidity range. 
The kaon and proton ratios  at mid-rapidity ($K^{-}/K^{+} \sim 0.85$, $\bar{p}/p \sim 0.49$) 
are lower than the corresponding ones measured at the top 
RHIC energy \cite{BRAHMSratios},
but they are still characteristic of a high degree of anti-matter to matter equilibration.
At forward rapidity we observe a decrease of the kaon ratio to a value of $\sim0.35$ at
$y=3.3$. 
Possible explanations include the competition between $\Lambda$ baryons and $K^{-}$ mesons for the available
strange quarks and associated production (e.g., $p+p \rightarrow p + \Lambda + K^{+}$) 
which increases the number of positive kaons. 
Both of these mechanisms depend on the net-baryon content and, consequently, 
lead to a decrease of the $K^{-}/K^{+}$ ratio at forward rapidity. 
The $\bar{p}/p$ ratio decreases significantly with rapidity, 
reaching $\bar{p}/p=0.077 \pm 0.004(stat.)$ at y=2.3 and $\bar{p}/p=0.022 \pm 0.001(stat.)$ at $y=3$.

\begin{figure}[htp]
\begin{center}
\includegraphics[width=1.0\columnwidth]{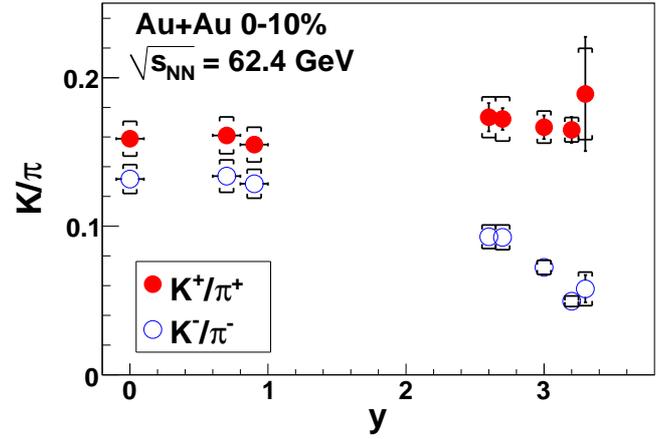}
\caption{(Color online) Rapidity dependence of the $K/\pi$ ratios in 0-10\% central Au+Au collisions at $\sqrt{s_{NN}}=62.4$~GeV. 
The error bars are statistical errors and the square brackets show the
systematic uncertainties due to the yield extrapolation at low $p_T$.}
\label{kpiRatios}
\end{center}
\end{figure}

In Fig.~\ref{kpiRatios} we show the rapidity dependence of the $K/\pi$ ratio. 
Because the rapidity intervals where the yields of the two species were extracted are not the same
at forward rapidity, 
we used a linear interpolation procedure between the closest covered points to obtain the meson
yields for additional points in rapidity.
We checked this procedure by assuming Gaussian rapidity distributions and
found very similar results.
The $K^{+}/\pi^{+}$ ratio was found to be $0.159\pm0.011$ at mid-rapidity and is almost 
constant as a function of rapidity. 
The $K^{-}/\pi^{-}$ ratio has a value of $0.13\pm0.01$ at mid-rapidity and shows a steep decrease
for $y>2.5$ with a value of $\sim0.05$ at $y=3.2$.
The different rapidity dependence of the positive and negative charge $K/\pi$ ratios 
is similar to that 
found in central Au+Au collisions at $\sqrt{s_{NN}}=200$~GeV  \cite{BRAHMSmeson}
but the difference between the two ratios is three times larger at y=3.

\begin{figure}[htp]
\begin{center}
\includegraphics[width=1.0\columnwidth]{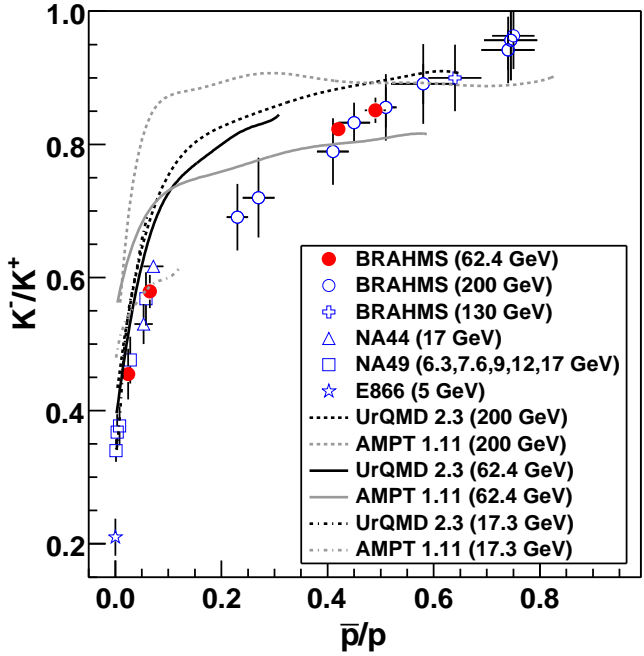}
\caption{(Color online) $K^{-}/K^{+}$ ratio dependence on the $\bar{p}/p$ ratio. 
The solid circles are from 0-10\% central Au+Au collisions at $\sqrt{s_{NN}}=62.4$~GeV obtained in the present work. 
The open symbols are BRAHMS data from ref. \cite{BRAHMSratios}
and lower energy data from \cite{SPSmeson, AGSratios, SPSratios, SPSkbark2030, SPSpbarp2030, NA44ratios}.
The error bars represent statistical and systematic errors.
The curves are calculations with UrQMD (black lines) and AMPT (gray lines) for central Au+Au collisions at $\sqrt{s_{NN}}$=200 GeV 
(dashed lines) and 62.4 GeV (solid lines) and for central Pb+Pb collisions at $\sqrt{s_{NN}}$=17.3 GeV (dot-dashed lines).
The curve for the UrQMD calculation at 17.3~GeV is very close to that at 62.4~GeV making it less visible.}
\label{kbarkVsPbarP}
\end{center}
\end{figure}

\begin{figure}[htp]
\begin{center}
\includegraphics[width=1.0\columnwidth]{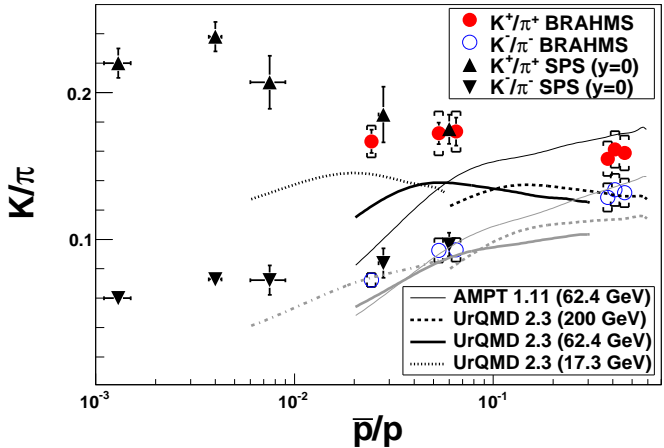}
\caption{(Color online) $K/\pi$ ratios as functions of the $\bar{p}/p$ ratio. 
The circles are data from 0-10\% central Au+Au collisions at $\sqrt{s_{NN}}=62.4$~GeV and the triangles
are results obtained at mid-rapidity in the SPS experiments at different energies \cite{SPSkpiEnergy}. 
The curves are UrQMD (thick lines) and AMPT (thin lines) calculations for central Au+Au collisions 
at $\sqrt{s_{NN}}=17.3, 62.4$ and $200$~GeV. The black curves are for the $K^{+}/\pi^{+}$ ratio
and gray curves are for the $K^{-}/\pi^{-}$ ratio.}
\label{kpiVsPbarp}
\end{center}
\end{figure}

In the following we use calculations with two microscopic transport models for comparison
with our data. Ultra-Relativistic Quantum Molecular Dynamics (UrQMD) \cite{UrQMD1, UrQMD2} 
is the extension of RQMD \cite{RQMD1, RQMD2} which was developed to describe physics at AGS and SPS energies.
At low energies, $\sqrt{s_{NN}}<5$~GeV, it describes the nuclear collisions in terms
of interactions between known hadrons and their resonances while at higher energies the 
dominant mechanism is through the color string excitation.
AMPT (A Multiphase Transport Model) \cite{AMPT1, AMPT2} is a microscopic model developed
for nucleus-nucleus collisions at RHIC and LHC energies. AMPT uses the HIJING model
for the initial space-time configuration of the partons and strings.
Both UrQMD and AMPT include a Lund-type string fragmentation procedure followed by
hadronic rescatterings. 
The models were run in minimum bias mode with a 0-10\% centrality cut placed on the distribution
of particles falling in the acceptance of our multiplicity detector ($|\eta| < 2.2$),
closely mimicking the experimental conditions.

Figure ~\ref{kbarkVsPbarP} shows the dependence of the $K^{-}/K^{+}$ ratio on the $\bar{p}/p$ ratio
in central nucleus-nucleus collisions at energies ranging from $\sqrt{s_{NN}}=5$ to
200~GeV. The data points obtained in this work and from the other RHIC energies 
are obtained in different rapidity slices, while the SPS and AGS points are obtained at
mid-rapidity. Comparing the low and high energy results, we observe a systematic dependence
of the $K^{-}/K^{+}$ ratio on the $\bar{p}/p$ ratio.
The calculations made with UrQMD(thick lines) and AMPT(thin lines)
for central nucleus-nucleus collisions at $\sqrt{s_{NN}}=200$, $62.4$ and $17.3$~GeV do not
reproduce quantitatively the dependence of
the $K^{-}/K^{+}$ ratio on the $\bar{p}/p$ ratio which seems to be universal over a large
energy range. However, UrQMD calculations at the three energies give similar $K^{-}/K^{+}$
ratios for a given $\bar{p}/p$ value. This is most likely due to the thermalization
of the system reached via secondary rescatterings, which includes formation, decay and regeneration
of many resonances \cite{urqmdThermalization}.

Figure ~\ref{kpiVsPbarp} shows the dependence of the $K/\pi$ ratios on the $\bar{p}/p$ ratio.
Our data points are obtained in different rapidity slices at the same energy, $\sqrt{s_{NN}}=62.4$~GeV, 
while the SPS points are obtained at mid-rapidity in central Pb+Pb collisions at 
$\sqrt{s_{NN}}=6.3, 7.6, 8.8, 12.3$ and $17.3$~GeV. 
In Au+Au collisions at $62.4$~GeV, the fragmentation peak is estimated to be in the interval $2.5<y<3.3$
\cite{BRAHMS62proton} with a net-proton density of $\sim30$ 
and $\bar{p}/p$ ratio values at forward rapidity, 
which are in the same range as those measured at mid-rapidity at $\sqrt{s_{NN}}=12.3$ and $17.3$~GeV \cite{SPSproton}.
As in the case of $K^{-}/K^{+}$ ratio, we observe a smooth dependence of the $K/\pi$ ratios on
the $\bar{p}/p$ ratio.
The curves show calculations for central Au+Au collisions
from UrQMD (thick lines) at $\sqrt{s_{NN}}=17.3, 62.4$ and 200~GeV and AMPT (thin lines) at $\sqrt{s_{NN}}=62.4$~GeV. 
None of the models reproduce quantitatively the dependence of the $K/\pi$ ratios on the $\bar{p}/p$ ratio
but qualitatively we observe that UrQMD reproduces the trend of the data.

\begin{figure}[htp]
\begin{center}
\includegraphics[width=1.0\columnwidth]{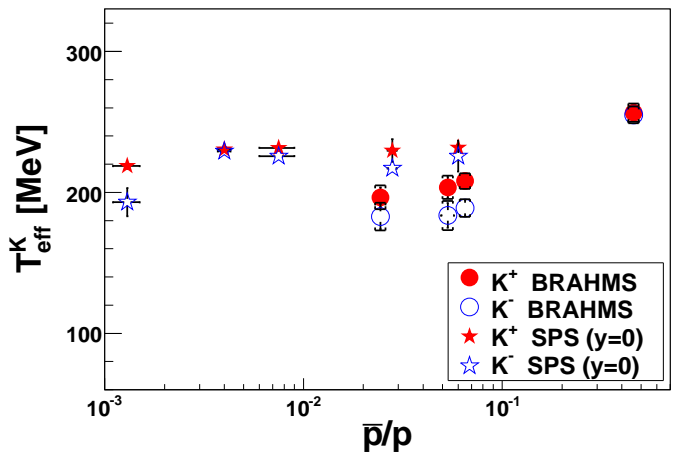}
\caption{(Color online) Inverse slopes ($T_{\mathrm{eff}}$) for kaons as a function of the $\bar{p}/p$ ratio. 
The BRAHMS points are from 0-10\% central Au+Au collisions at $\sqrt{s_{NN}}=62.4$~GeV (this analysis)
at different rapidities while the SPS points are from mid-rapidity 
at different energies \cite{SPSmeson, SPSkbark2030}.
The error bars represent statistical errors, while the square brackets show systematic errors.}
\label{kaonTeff}
\end{center}
\end{figure}

Figure ~\ref{kaonTeff} displays the inverse slope parameters for kaons as a function
of the $\bar{p}/p$ ratio. The inverse slope for the highest ratio (0.45) represents the result from a simultaneous fit to the spectra in $0<y<1$.
Also shown are the mid-rapidity results from SPS.
The inverse slopes of the spectra for positive and negative kaons measured at forward rapidity
in our dataset are $\sim20$ MeV smaller than the ones measured at mid-rapidity in SPS experiments 
at the same $\bar{p}/p$ ratio.
Even though the values are consistent within error bars, the difference suggests that the 
transverse momenta are not governed by the baryo-chemical potential.
The mid-rapidity particle densities for negative pions
at SPS at $\sqrt{s_{NN}}=17.3$ and 12.3~GeV are approximately two times larger \cite{SPSmeson} than the negative pion densities at the forward rapidities at $\sqrt{s_{NN}}=$62.4~GeV. 
Smaller radial flow from the less dense local system at 62.4 GeV could be the cause of the small difference between inverse slopes.
The inverse slopes obtained at forward rapidity are higher for positive kaons
than for the negative kaons. This is also observed at mid-rapidity at SPS energies.

In a chemical analysis, the $\bar{p}/p$ ratio has an approximate correspondence with the baryo-chemical 
potential through the formula $\bar{p}/p = \mathrm{exp}(-2\mu_B/T)$ for a given freeze-out temperature $T$.
Hence if $T$ is the same in the two cases, this would imply that the local system formed at high rapidity at RHIC (62.4 GeV) 
is chemically equivalent with the system formed at the two highest SPS energies at mid-rapidity, both being
controlled by the baryo-chemical potential.
The strangeness chemical potential is fixed by the baryo-chemical potential and
chemical freeze-out temperature provided that the local net strangeness vanishes.
A rapidity dependent baryo-chemical potential $\mu_B$ has been suggested in thermal 
models \cite{Becattini, broniowski, stiles}.
The observed dependence of the chemical freeze-out composition in different colliding systems and
different rapidities only on the baryo-chemical
potential and temperature supports the idea of local chemical equilibration and the existence of a
universal ($T-\mu_{B}$) freeze-out line. 

\section*{Summary}
In summary, we have measured the transverse momentum spectra and inclusive invariant yields of charged
pions and kaons in central Au+Au reactions at 62.4 GeV. The anti-particle/particle ratios for kaons and protons show a 
steep decrease at forward rapidity. The charge dependence of the $K/\pi$ ratio at forward rapidities
is understood in the framework of microscopic models (\textit{i.e.} UrQMD) 
as resulting from  the associated production
in a baryon rich medium. The production mechanisms enhance the fraction of $s$ quarks ending up in hyperons,
thus depleting the $K^{-}$ yield.
We observe in central nucleus-nucleus collisions a common dependence of the particle ratios 
($K/\pi$, $K^{-}/K^{+}$) and kaon spectra inverse slopes on $\bar{p}/p$, which reflects the baryo-chemical
potential, whether
measured for different energies at mid-rapidity at SPS, or at different rapidities at 
$\sqrt{s_{NN}}=62.4$~GeV. 
This is consistent with a picture where
in nucleus-nucleus collisions, at a given energy, the local fireballs formed
at different rapidities freeze out on a $T-\mu_{B}$ line which coincides with
the $T-\mu_{B}$ freeze-out line previously observed in a wide energy range but only at mid-rapidity.
The baryo-chemical potential interval covered at forward rapidity extends to high values 
and is equivalent to the two highest SPS energies, but not quite high enough to probe the horn 
structure in the $K^{+}/\pi^{+}$ ratio excitation function.
In the calculations made with UrQMD and AMPT models, $K/\pi$ ratios are not well
reproduced for large baryo-chemical potential found either at high rapidity or mid-rapidity
at lower energies although UrQMD seems to reproduce qualitatively the trend of the data. 
Also, the universal dependence of the $K^{-}/K^{+}$ on the $\bar{p}/p$ ratio
is not well explained quantitatively by these microscopic transport models.
However the UrQMD model seems to reproduce qualitatively very well this dependence.

\section*{Acknowledgements}
This work was supported by the Division of Nuclear Physics of the
Office of Science of the U.S. Department of Energy under contracts
DE-AC02-98-CH10886, DE-FG03-93-ER40773, DE-FG03-96-ER40981, and
DE-FG02-99-ER41121, the Danish Natural Science Research Council,
the Research Council of Norway, the Polish Ministry of Science and 
Higher Education (Contract no 1248/B/H03/2009/36), and the Romanian
Ministry of Education and Research (5003/1999, 6077/2000). We
thank the staff of the Collider-Accelerator Division at BNL and the RHIC Computing Facility for
their support to the experiment.



\end{document}